\documentclass[lettersize,journal]{IEEEtran}

\usepackage{amsmath,amssymb,amsthm,fixmath}
\usepackage{bm}
\usepackage{cite}
\usepackage{hyperref}
\hypersetup{
    colorlinks=true,
    linkcolor={black},
    citecolor={black},
    urlcolor={black}
    }
\usepackage{pgfplots}
\pgfplotsset{compat=1.18}
\usepackage{algorithm}
\usepackage{algorithmic}
\usepackage{tabularx}
\usepackage{cleveref}
\usepackage{xcolor}
\usepackage{pgfplots}
\usepackage{tikz}
\usetikzlibrary{intersections,pgfplots.fillbetween,patterns,spy,shapes.geometric,calc,positioning}
\usetikzlibrary{arrows,decorations.pathmorphing,backgrounds,fit,matrix,fillbetween}
\usetikzlibrary{arrows.meta}
\usetikzlibrary{decorations.pathreplacing,decorations.markings}
\usepackage{tumcolor}
\usepackage{balance}

\usepackage[acronym,shortcuts]{glossaries}
\newacronym{gmm}{GMM}{Gaussian mixture model}
\newacronym{cme}{CME}{conditional mean estimator}
\newacronym{pdf}{PDF}{probability density function}
\newacronym{mmse}{MMSE}{minimum mean square error}
\newacronym{rv}{RV}{random variable}
\newacronym{snr}{SNR}{signal-to-noise ratio}
\newacronym{mse}{MSE}{mean square error}
\newacronym{awgn}{AWGN}{additive white Gaussian noise}
\newacronym{simo}{SIMO}{single-input multiple-output}
\newacronym{adc}{ADC}{analog-to-digital converter}

\usepackage{siunitx} 

\newtheorem{theorem}{Theorem}

\newtheorem{corollary}{Corollary}

\newtheorem{remark}{Remark}

\newcommand{\B}[1]{{\bm{#1}}}
\newcommand{\dx}[1]{\operatorname{d}\hspace{-1.5pt}#1}
\newcommand{\op}[1]{{\operatorname{#1}}}
\newcommand{\h}{^{\operatorname{H}}}
\newcommand{\T}{^{\operatorname{T}}}

\newcommand{\inv}{^{-1}}

\newcommand*{\C}{\mathbb{C}}

\DeclareMathOperator{\eye}{\mathbf{I}}

\DeclareMathOperator{\vect}{vec}

\DeclareMathOperator{\diag}{diag}
\DeclareMathOperator{\sign}{sign}
\DeclareMathOperator{\E}{\mathbb{E}}

\DeclareMathOperator{\NC}{\mathcal{N}_{\mathbb{C}}}

\newcommand{\hhat}{\hat{\B h}}

\pgfplotsset{
  every axis legend/.append style={
    font=\scriptsize 
  }
}

\pgfplotsset{
  every axis title/.append style={
    font=\scriptsize 
  }
}

\pgfplotsset{every axis/.append style={
                    label style={font=\scriptsize},
                    tick label style={font=\scriptsize}  
                    }}

\glsdisablehyper

\begin{document}
    \bstctlcite{IEEEexample:BSTcontrol}
    \title{Linear and Nonlinear MMSE Estimation in One-Bit Quantized Systems under a Gaussian Mixture Prior}
	\author{Benedikt Fesl,~\IEEEmembership{Graduate Student Member,~IEEE} and Wolfgang Utschick,~\IEEEmembership{Fellow,~IEEE}
		\thanks{
			The authors are with Chair of Signal Processing, Technical University of Munich, Munich, Germany  (e-mail: benedikt.fesl@tum.de; utschick@tum.de).
		}
	}
	
	\maketitle

	\begin{abstract}
        We present new fundamental results for the \ac{mse}-optimal \ac{cme} in one-bit quantized systems for a \ac{gmm} distributed signal of interest, possibly corrupted by \ac{awgn}. 
        We first derive novel closed-form analytic expressions for the Bussgang estimator, the well-known linear \ac{mmse} estimator in quantized systems.
        Afterward, closed-form analytic expressions for the \ac{cme} in special cases are presented, revealing that the optimal estimator is linear in the one-bit quantized observation, opposite to higher resolution cases. 
        Through a comparison to the recently studied Gaussian case, we establish a novel \ac{mse} inequality and show that that the signal of interest is correlated with the auxiliary quantization noise.
        We extend our analysis to multiple observation scenarios, examining the \ac{mse}-optimal transmit sequence and conducting an asymptotic analysis, yielding analytic expressions for the \ac{mse} and its limit.
        These contributions have broad impact for the analysis and design of various signal processing applications.
	\end{abstract}
	

\begin{figure}[b]
\onecolumn
\centering
\copyright \scriptsize{This work has been submitted to the IEEE for possible publication. Copyright may be transferred without notice, after which this version may no longer be accessible.}
\vspace{-1.3cm}
\twocolumn
\end{figure}

\section{Introduction}
\IEEEPARstart{B}{ayesian} estimators are a cornerstone in classical estimation theory, affecting many signal processing applications. In particular, the \ac{cme} as the optimal estimator for all Bregman loss functions, with the \ac{mse} as the most prominent representative \cite{1459065}, is of great importance.
This has led to the analysis of the \ac{cme} and its properties under various conditions, e.g., in an \ac{awgn} channel \cite{5730572} or with different noise models \cite{8989246}. In particular, the cases where the \ac{cme} is linear are of great interest due to the practical implications \cite{6157621}.

Despite its great importance in signal processing, the \ac{cme} lacks theoretical understanding in quantized systems, which, e.g., occur in the modeling of \acp{adc}, imposing a nonlinear inverse problem. Natural fields of application are, e.g, lossy compression \cite{8416707}, wireless sensor networks \cite{8645383}, audio coding \cite{audio_coding}, control theory \cite{quantized_meas}, positioning \cite{localization_stein}, or channel estimation \cite{Ivrlac2007,7931630}.
Recently, the \ac{cme} was examined in case of one-bit quantization in a jointly Gaussian setting \cite{fesl23cme1bit}, \cite{ding2024optimal}; it was shown that the \ac{cme} is linear in the quantized observation in many special cases, although it necessitates an elaborate numerical evaluation in general.

A viable alternative, in general, is the linear \ac{mmse} estimator, which can be derived via the Bussgang decomposition \cite{9307295,7931630,9046300}, motivated by Bussgang's theorem \cite{Bussgang}, or, alternatively, the additive quantization noise model \cite{aqnm}. Moreover, the statistically equivalent linear model via the Bussgang decomposition allows for theoretical system analysis, e.g., spectral efficiency \cite{7931630}, capacity \cite{ISIT2012_mezghani_nossek_ieee}, nonideal hardware effects \cite{6891254,8501941}, or nonlinear system characterization \cite{837046,871400,7373634}. However, similar to the analysis of the \ac{cme}, the Bussgang estimator was mainly investigated for the case of a zero-mean Gaussian distributed signal, allowing for closed-form solutions of the Bussgang gain \cite[Sec. 9.2]{Papoulis1965ProbabilityRV} and the covariance matrix of the quantized observation via the \textit{arcsine law} \cite{1446497,272490}.

A natural generalization of the zero-mean Gaussian case is the zero-mean \ac{gmm}, which covers a wide class of \acp{pdf} that can be reasonably approximated, especially in wireless communications~\cite{10454252,boeck2024statistical}.
This has motivated the analysis of the Bussgang gain for \ac{gmm} distributed signals \cite{1583931,banelli2013nonlinear}. However, the Bussgang decomposition has not been fully investigated for the general multivariate case for one-bit quantization. More importantly, the linear \ac{mmse} and the \ac{cme} for a \ac{gmm} prior are not investigated thus far.

The \textit{contributions} of this letter are as follows: 
We generalize the analytic expressions for the Bussgang gain and the arcsine law from the Gaussian case to the general \ac{gmm} case, which is important for the evaluation of the linear \ac{mmse} estimator. Afterward, we study the \ac{cme} estimator in different special cases.  
We derive a novel closed-form solution for the \ac{cme} in the univariate case, which turns out to be linear in the observation and thus equal to the Bussgang estimator. This allows for finding an analytic expression of the cross-correlation between the signal of interest and the auxiliary quantization noise from the Bussgang decomposition, which is vanishing in both the low and high \ac{snr} regimes or in the degenerate Gaussian case. 
Furthermore, we derive a novel \ac{mse} inequality, revealing that the \ac{gmm} distribution leads to a consistently higher \ac{mse} than the Gaussian distribution under a fixed global variance constraint. 

Subsequently, we investigate a multiple observation scenario, where the \ac{mse}-optimal observation sequence and two equivalent expressions of the \ac{cme} for the noiseless case are derived. Subsequent to finding analytic expressions of the \ac{mse} and its limit, the \ac{mse} inequality is shown to also hold for this case. All theoretical results are validated with numerical experiments. In addition, more general cases are evaluated, highlighting the strong impact of \textit{stochastic resonance}.

\section{System Model}\label{sec:problem_formulation}
We consider the generic system equation $\B R = Q(\B Y) =  Q(\B h\B a\T + \B N) \in \mathbb{C}^{N\times M}$ where $\B R=[\B r_1, \B r_2, \dots,\B r_M]$ contains $M$ quantized observations of the vector of interest $\B h \in\mathbb{C}^N$ with the known vector $\B a \in\mathbb{C}^M$ that fulfills the power constraint $\|\B a\|_2^2= M$. 
Let the vector $\B h \sim p(\B h)$ be a zero-mean \ac{gmm} \ac{rv}, i.e., 
\begin{align}
    p(\B h) = \sum_{k=1}^K p_k\mathcal{N}_{\mathbb{C}}(\B h; \B 0, \B C_k)
\end{align}
with the global covariance matrix $\B C_{\B h} = \sum_{k=1}^K p_k \B C_k$.
Further, $\B N = [\B n_1, \dots, \B n_M]$ where $\B n_i\sim\mathcal{N}_{\mathbb{C}}(\B 0, \eta^2\eye)$ is \ac{awgn} and $Q(\cdot) = \frac{1}{\sqrt{2}} \left(\sign(\Re(\cdot)) + \op j \sign(\Im(\cdot))\right)$ is the complex-valued one-bit quantization function,
which is applied element-wise to the input vector/matrix. The system model can be equivalently described in its (column-wise) vectorized form as
\begin{equation}
	\B r = Q(\B y) = Q(\B A \B h  + \B n) \in \mathbb{C}^{NM}
	\label{eq:system_vec}
\end{equation}
with $\B A = \B a \otimes \eye$, $\B r = \vect(\B R)$, $\B y = \vect(\B Y)$, and $\B n = \vect(\B N)$.

\section{The Bussgang Estimator}\label{sec:Bussgang}
In the context of quantization, the linear \ac{mmse} estimator is referred to as the Bussgang estimator \cite{7931630}, as it is motivated by Bussgang's theorem~\cite{Bussgang}. In particular, the Bussgang decomposition implies that the system \eqref{eq:system_vec} can be linearized as the statistically equivalent model
\begin{align}\label{eq:Bussgang_decomp}
    \B r = Q(\B y) = \B B \B y + \B q,
\end{align}
where $\B B$ is the Bussgang gain, enforcing that $\B q$ is uncorrelated (but not independent) of $\B y$. The Bussgang estimator reads as
\begin{align}\label{eq:lmmse}
    \hhat_{\text{LMMSE}} = \B C_{\B h\B r} \B C_{\B r}\inv \B r
    = (\B C_{\B h} \B A\h \B B\h + \B C_{\B h\B q} ) \B C_{\B r}\inv \B r.
\end{align}
In the case of jointly zero-mean Gaussian quantizer input, it is well-known that $\B B$ and $\B C_{\B r}$ can be computed in closed-form, and $\B C_{\B h\B q} = \B 0$, cf. \cite{7931630}, \cite[Sec. 9.2]{Papoulis1965ProbabilityRV}.
Although the zero-mean \ac{gmm} is a natural generalization, covering a much larger class of \acp{pdf} that can be approximated, the expressions for $\B C_{\B h\B r}$, $\B B$, and $\B C_{\B r}$ in case of one-bit quantization are not fully investigated thus far. This motivates the derivation of these expressions in the following. 

\begin{theorem}\label{theorem:lmmse}
The involved quantities for the linear \ac{mmse} estimator \eqref{eq:lmmse} are computed as
\begin{align}
    \B B &= \sqrt{\frac{2}{\pi}}\sum_{k=1}^K p_k \diag(\B C_{\B y|k})^{-\frac{1}{2}}\B C_{\B y|k}\B C_{\B y}\inv,
        \label{eq:Bussgang_gain_general}
    \\
    \B C_{\B h \B r} &= \sqrt{\frac{2}{\pi}}\sum_{k=1}^K p_k \B C_{k} \B A\h \diag(\B C_{\B y|k})^{-\frac{1}{2}},
    \label{eq:Chr}
    \\
    \B C_{\B r} &= \frac{2}{\pi} \sum_{k=1}^K p_k (\sin\inv(\Re(\bar{\B C}_{\B y|k})) + \op j \sin\inv(\Im(\bar{\B C}_{\B y|k})))
    \label{eq:arcsin}
\end{align}
with $\bar{\B C}_{\B y|k} = \diag(\B C_{\B y|k})^{-\frac{1}{2}} \B C_{\B y|k} \diag(\B C_{\B y|k})^{-\frac{1}{2}}$ and $\B C_{\B y} = \sum_{k=1}^K p_k \B C_{\B y|k}$ where $\B C_{\B y | k} = \B A \B C_k\B A\h + \eta^2\eye$.
\end{theorem}
\textit{Proof:} See Appendix \ref{app:theorem_lmmse}.

\begin{remark}
    The Bussgang gain \eqref{eq:Bussgang_gain_general} is in accordance with the findings in \cite{banelli2013nonlinear}; however, the authors only discuss the univariate case, and one-bit quantization is not analyzed. 
    In contrast to the Gaussian case, the Bussgang gain \eqref{eq:Bussgang_gain_general} is not a diagonal matrix in general, which aligns with the statement in \cite{9307295}.
    The expression \eqref{eq:arcsin} can be interpreted as a weighted version of the arcsine law \cite{1446497,272490}.
    Since \acp{gmm} are universal approximators \cite{NgNgChMc20}, a straightforward application of the above results is to approximate an unknown density with a \ac{gmm}, allowing to compute the linear \ac{mmse} estimator. A similar approach was adopted in \cite{10454252}.
\end{remark}

\begin{corollary}\label{cor:corr}
    Based on the results of Theorem \ref{theorem:lmmse}, the cross-covariance matrix $\B C_{\B h\B q} = \B C_{\B h\B r} - \B C_{\B h} \B A\h \B B\h$ of the signal of interest $\B h$ and the auxiliary quantization noise $\B q$ in \eqref{eq:Bussgang_decomp} is
    \begin{align}
    \begin{aligned}
        \B C_{\B h\B q} = \sqrt{\frac{2}{\pi}} \sum_{k=1}^K p_k &\left(\B C_k \B A\h \diag(\B C_{\B y|k})^{-\frac{1}{2}}
        \right.
        \\
        &\left.- \B C_{\B h}\B A\h \diag(\B C_{\B y|k})^{-\frac{1}{2}} \B C_{\B y|k}\B C_{\B y}\inv \right),
    \end{aligned}
    \end{align}
    contrary to the Gaussian case, where $\B C_{\B h\B q} = \B 0$~\cite{7931630}.
\end{corollary}

After deriving the linear \ac{mmse} estimator for the general case, we investigate the \ac{cme} in the following, where we particularly discuss special cases in which it is linear.

\section{The Conditional Mean Estimator}
In the general case, the \ac{cme} is not analytically tractable, necessitating a numeric approach to solve the involved integral expressions.
However, when rewriting the \ac{cme} as
\begin{align}
    \E[\B h | \B r] &= \sum_{k=1}^K p(k | \B r) \E[\B h | \B r, k]
    \label{eq:cme_first_step}
    \\
    &= \sum_{k=1}^K \frac{p_k}{\sum_{i=1}^K p_i p(\B r | i)}\int \B h p(\B h | k) p(\B r | \B h) \dx \B h,
    \label{eq:cme_convinient}
\end{align}
utilizing the law of total expectation and Bayes' rule, all involved densities are conditioned on a \ac{gmm} component. This allows for effectively treating them as in the Gaussian case, directly enabling simplified numerical evaluations discussed in \cite{fesl23cme1bit,ding2024optimal}, whose discussion is left out due to space limitations.
However, we derive novel closed-form analytic solutions of the \ac{cme} for special cases in the following, accompanied by comparisons to the Gaussian case analyzed in \cite{fesl23cme1bit,ding2024optimal}.

\subsection{Univariate Case with a Single Observation}
We consider the case of a scalar system $r = Q(h + n)$ with $h\sim \sum_{k=1}^Kp_k\mathcal{N}_\C(0, \sigma_k^2)$ and $n\sim\NC(0, \eta^2)$.

\begin{theorem}\label{theorem:gmm_cme_1bit_univariate}
The \ac{cme} for the scalar system is computed as
    \begin{align}
         \E[h | r] = \sqrt{\frac{2}{\pi}} \sum_{k=1}^K p_k \frac{\sigma_k^2}{\sqrt{\sigma_k^2 + \eta^2}}r.
         \label{eq:cme_gmm_1bit_univariate}
    \end{align}
\end{theorem}
\textit{Proof:} See Appendix \ref{app:theorem_cme_univariate}.

\begin{remark}
    Remarkably, in contrast to the high-resolution case, the \ac{cme} is linear in the quantized observation, i.e., the optimal estimator becomes linear through the specific nonlinearity of the quantization process. Furthermore, the jointly Gaussian case is not unique for the \ac{cme} to be linear, in contrast to linear \ac{awgn} channels \cite{6157621}.
    The result can be immediately extended to a multivariate zero-mean \ac{gmm} with diagonal covariances.
\end{remark}

The \ac{mse} of the \ac{cme} is then computed to
\begin{align}
    \text{MSE}_{\textup{GMM}} = \sigma_{\text{glob}}^2 - \frac{2}{\pi} \left(\sum_{k=1}^K p_k \frac{\sigma_k^2}{\sqrt{\sigma_k^2 + \eta^2}} \right)^2
    \label{eq:mse_univariate}
\end{align}
where $\sigma^2_{\text{glob}} = \sum_{k=1}^K p_k \sigma_k^2$ is the global variance.

An interesting analysis is a comparison to the case of a Gaussian distributed \ac{rv} with the same global variance, i.e., $h\sim \NC(0, \sigma_{\text{glob}}^2)$, for which the closed-form \ac{mse} of the corresponding \ac{cme} is given as, cf. \cite{fesl23cme1bit},
\begin{align}
    \textup{MSE}_{\textup{Gauss}} = \sigma_{\text{glob}}^2 - \frac{2}{\pi}\frac{\sigma_{\text{glob}}^4}{\sigma_{\text{glob}}^2 + \eta^2}.
    \label{eq:mse_gauss}
\end{align}
This allows to compare the estimation performance of the \acp{cme} when changing the distribution of the \ac{rv} of interest from a Gaussian to a \ac{gmm} while keeping the global variance fixed. We note that both estimators are optimal with respect to the considered distribution.
\begin{theorem}
    \label{theorem:mse_inequality}
    For the considered scalar system, under a fixed global variance $\sigma_{\textup{glob}}^2$, it holds for all \acp{snr} that
    \begin{align}
        \textup{MSE}_{\textup{Gauss}} \leq \textup{MSE}_{\textup{GMM}}.
        \label{eq:theorem_mse_inequality}
    \end{align}
\end{theorem}
\textit{Proof:} See Appendix \ref{app:theorem_mse_inequality}.

\begin{remark}\label{remark:mse_inequality}
    In the high \ac{snr} regime, we get 
\begin{align}
    \lim_{\eta^2\to 0} \textup{MSE}_{\textup{GMM}} = \sigma_{\textup{glob}}^2 - \frac{2}{\pi} 
    \bar{\sigma}^2
    \label{eq:mse_univariate_noiseless}
\end{align}
with $\bar{\sigma} = \sum_{k=1}^K p_k \sigma_k$,
and the inequality \eqref{eq:theorem_mse_inequality} directly follows from the weighted Cauchy-Schwarz inequality
\begin{align}
    \bar{\sigma}^2 \leq \sigma_{\textup{glob}}^2
    \label{eq:CS_inequality}
\end{align}
with equality if and only if $\sigma_k^2 = \sigma^2$ for all $k=1,\dots,K$, i.e., when the \ac{gmm} degenerates to a Gaussian.
The observation that a \ac{gmm} distribution leads to a strictly higher \ac{mmse} than the Gaussian distribution under a fixed global variance constraint is not stated in the literature so far.
\end{remark}

Due to the linearity of the \ac{cme}, it is equal to the Bussgang estimator, representing the linear \ac{mmse}, cf. \Cref{sec:Bussgang}. Using the result of Corollary \ref{cor:corr}, we 
can further compute 
\begin{align}
    \E[hq^*] =\sqrt{\frac{2}{\pi}} \sum_{k=1}^K p_k \left( \frac{\sigma_k^2}{\sqrt{\sigma_k^2 + \eta^2}} - \frac{\sigma_{\text{glob}}^2 \sqrt{\sigma_k^2 + \eta^2}}{\sigma_{\text{glob}}^2 + \eta^2}\right),
    \label{eq:correlation_quant_noise}
\end{align}
which is in contrast to the Gaussian case where $\E[hq^*]~=~0$ \cite[Appendix A]{7931630}. 
Moreover, the correlation $\E[hq^*]$ vanishes in both the low and high \ac{snr} regime, i.e.,
\begin{align}
\label{eq:corr_vanish}
    \lim_{\eta\to\infty} \E[hq^*] = \lim_{\eta\to0} \E[hq^*] = 0.
\end{align}

\subsection{Univariate Noiseless Case with Multiple Observations}
\label{sec:multiple_observations}

We consider the noiseless case with multiple pilot observations $\B r = Q(\B ah)$, for which the closed-form \ac{cme} together with the optimal pilot sequence was derived in \cite{fesl23cme1bit} for the Gaussian case. First, we show that the pilot sequence for the Gaussian case is also \ac{mse}-optimal for the \ac{gmm} case.

\begin{theorem}\label{theorem:pilot_sequence}
    The \ac{mse}-optimal pilot sequence for the considered system contains equidistant phase shifts $\psi_m = \frac{\pi(m-1)}{2M}$ for all $m=1,\dots,M$, such that $[\B a]_m = \exp(\op j \psi_m)$.
\end{theorem}

\text{Proof:} See Appendix \ref{app:pilot_sequence}.

\begin{remark}
    In contrast to the Gaussian case, the amplitudes of the \ac{gmm} are not Rayleigh distributed, which does not impact the design of the optimal pilot sequence since the amplitude information is lost through the one-bit quantization, and the circular symmetry property is not affected \cite{340781}.
\end{remark}

\begin{theorem}\label{theorem:cme_multiple}
    The \ac{cme} for the considered system has the two equivalent expressions
    \begin{align}
        \E[h | \B r] &= \sqrt{\frac{2}{\pi}} \sum_{k=1}^K p_k \sigma_k \B a\h \B C_{\B r}\inv \B r
        \label{eq:cme_linear}
        \\
        &= \sum_{k=1}^K p_k \frac{2M\sigma_k}{\sqrt{\pi}}\sin\left(\frac{\pi}{4M}\right)\exp\left(\op j \varphi(\B r)\right)
        \label{eq:cme_nonlinear}
    \end{align}
    where $\B C_{\B r}\inv$ is equivalent to the analytic expression that solely depends on the number of pilots from the Gaussian case \cite[Lemma 1]{fesl23cme1bit}, and $\varphi(\B r) = \angle(\frac{1}{M} \sum_{m=1}^M [\B r]_m) - \frac{(M-1)\pi}{4M}$ \cite{fesl23cme1bit}.
\end{theorem}
\textit{Proof:} See Appendix \ref{app:cme_multiple}.

\begin{remark}
    The result of Theorem \ref{theorem:cme_multiple} is interesting since it leads to two equivalent formulations of the \ac{cme}, one being linear and one being nonlinear in the observation. This allows for different but equivalent implementations of the optimal estimator in a practical system. Moreover, the expression \eqref{eq:cme_nonlinear} allows for a simplified formulation of the closed-form analytic \ac{mse} in the following.
\end{remark}

The \ac{mse} of the \ac{cme} \eqref{eq:cme_nonlinear} is computed to
\begin{align}\label{eq:mse_pilots}
    \text{MSE}_{\text{GMM}}
    &= \sigma_{\text{glob}}^2 - \frac{4M^2}{\pi}\sin^2\left(\frac{\pi}{4M}\right)\bar{\sigma}^2.
\end{align}
Thus, we get in the limit of infinitely many pilots, cf. \cite{fesl23cme1bit},
\begin{align}\label{eq:mse_pilots_limit}
    \lim_{M\to \infty} \text{MSE}_{\text{GMM}} = \sigma_{\text{glob}}^2 - \frac{\pi}{4}\bar{\sigma}^2.
\end{align}
Observing the \ac{mse} expression for the Gaussian case in \cite{fesl23cme1bit}, under a fixed global variance $\sigma_{\text{glob}}^2$, we directly see by the weighted Cauchy-Schwarz inequality \eqref{eq:CS_inequality} that 
\begin{align}\label{eq:mse_inequality_pilots}
    \text{MSE}_{\text{Gauss}} \leq \text{MSE}_{\text{GMM}}
\end{align}
holds for all numbers of observations $M$, generalizing the result from \eqref{eq:theorem_mse_inequality}.

\section{Numerical Results}
For each simulation, we draw $10{,}000$ samples from a \ac{gmm} for estimating the normalized \ac{mse} $\E[\|\B h - \hhat\|_2^2] / \E[\|\B h\|_2^2]$. 

\begin{figure}[t]
    \centering
    \includegraphics{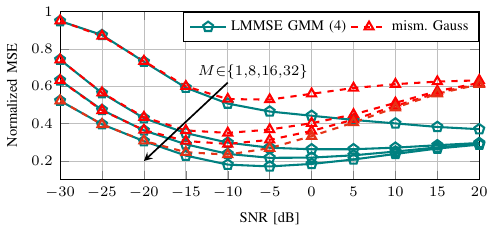}
	\caption{Comparison of the linear \ac{mmse} estimator with the suboptimal estimator assuming a Gaussian prior for $N=64$ and $M\in\{1,8,16,32\}$.}
  \label{fig:lmmse_comp}
\end{figure}

In Fig. \ref{fig:lmmse_comp}, we choose a ground-truth \ac{gmm} with $K=2$ components with the weights $p_{1} = 0.8$ and $p_{2} = 0.2$ and $N=64$-dimensional randomly chosen covariances following to the procedure in \cite{fesl23cme1bit}, which are afterward scaled with a factor of $0.1$ and $10$ for $k=1$ and $k=2$, respectively. 
This resembles a simple \ac{gmm} where the differences to the Gaussian case are evident.
We compare the linear \ac{mmse} estimator \eqref{eq:lmmse} with the quantities derived in Theorem \ref{theorem:lmmse} to the suboptimal linear Bussgang estimator assuming a Gaussian prior $\NC(\B 0, \B C_{\B h})$ (mism. Gauss) for $M\in \{1,8,16,32\}$ with the pilot sequence from Theorem \ref{theorem:pilot_sequence}. The Gaussian approximation is tight in the low \ac{snr} regime but shows a considerable gap for medium and high \acp{snr}, highlighting the importance of the novel derived linear \ac{mmse} estimator for the \ac{gmm}~case.

In the following, we consider the same \ac{gmm} but for $N=1$, where we choose $\sigma_1^2 = 0.1$ and $\sigma_2^2 = 10$, which are afterward normalized such that $\sigma_{\text{glob}}^2 = 1$. This ensures that the inequality \eqref{eq:CS_inequality} has a non-negligible gap.

\begin{figure}[t]
    \centering
    \includegraphics{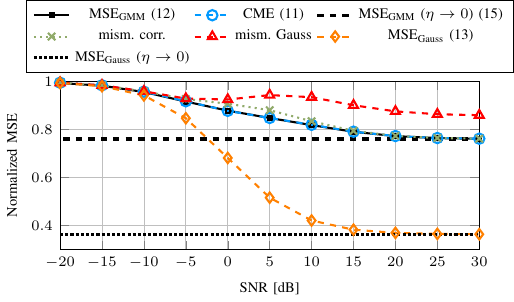}
	\caption{\ac{mse} results of the \ac{cme} from Theorem \ref{theorem:gmm_cme_1bit_univariate} for the univariate case $r = Q(h+n)$ in comparison with the Gaussian case, validating Theorem \ref{theorem:mse_inequality}.}
    \label{fig:univariate_single}
\end{figure}

We first verify the result of Theorem \ref{theorem:gmm_cme_1bit_univariate} for the univariate case with $N=M=1$ in Fig. \ref{fig:univariate_single}. It can be seen that the analytic \ac{mse} expression \eqref{eq:mse_univariate} is on par with the evaluation of the \ac{cme} \eqref{eq:cme_gmm_1bit_univariate}, converging to the noiseless case \eqref{eq:mse_univariate_noiseless}. Additionally, we have evaluated the estimator where the cross-correlation $\E[hq^*]$ is neglected (mism. corr.), which shows a performance loss in the medium \ac{snr} regime, being in accordance with \eqref{eq:corr_vanish}; the estimator that erroneously assumes a Gaussian distributed input, evaluating the \ac{cme} from \cite{fesl23cme1bit} (mism. Gauss) deteriorates from the \ac{cme} with a considerable gap. The \ac{cme} when the distribution is changed to a Gaussian and its limit show a clearly lower \ac{mse} over the whole \ac{snr} range, validating Theorem~\ref{theorem:mse_inequality} and Remark~\ref{remark:mse_inequality}.

\begin{figure}[t]
    \centering
    \includegraphics{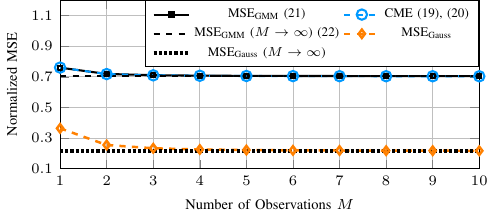}
	\caption{Performance of the \ac{cme} from Theorem \ref{theorem:cme_multiple} for the \ac{mse}-optimal pilot sequence from Theorem \ref{theorem:pilot_sequence} in comparison to the Gaussian case.}
 \label{fig:pilots}
\end{figure}

Fig. \ref{fig:pilots} assesses the \ac{cme} from Theorem \ref{theorem:cme_multiple} for the \ac{mse}-optimal pilot sequence in Theorem \ref{theorem:pilot_sequence} in the noiseless case. It can be observed that the limit is achieved already with a few observations, similar to the Gaussian case \cite{fesl23cme1bit}, which yields a lower \ac{mse} for all observations, cf. \eqref{eq:mse_inequality_pilots}.

\begin{figure}[t]
    \centering
    \includegraphics{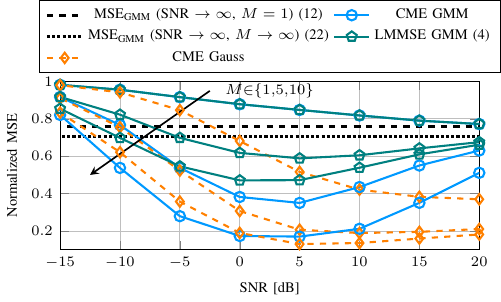}
	\caption{Comparison of the \ac{cme} with the \ac{mse}-optimal transmit sequence to the Gaussian case for $\B r = Q(\B a h + \B n)$ with $M\in\{1,5,10\}$.}
  \label{fig:pilots_noisy2}
\end{figure}

Finally, Fig. \ref{fig:pilots_noisy2} evaluates the \ac{cme} for the noisy case with multiple observations, i.e., $\B r = Q(\B a h +\B n)$, which has no analytic expression and is computed numerically via the algorithm from \cite{Genz1992}, implemented in \cite{2020SciPy}, cf. \cite{fesl23cme1bit}. It can be seen that the \ac{mse} limit for infinitely many observations without \ac{awgn} is drastically outperformed with only a few observations and finite \acp{snr}. This behavior is attributed to the fundamental effect of \textit{stoachstic resonance} \cite{mcdonnell_stocks_pearce_abbott_2008}, where noise can improve the performance in a quantized system. 
In comparison to the Gaussian case with the same global variance $\sigma_{\text{glob}}^2 = 1$ (CME Gauss), the stochastic resonance effect seems to be more pronounced, and the \ac{mse} inequality does not hold anymore, especially with more observations and in the low \ac{snr} regime. Moreover, the sub-optimal low-complexity linear \ac{mmse} estimator \eqref{eq:lmmse} (LMMSE GMM) degrades from the \ac{cme}, especially for higher numbers of observations. 

Further results are shown in Appendix \ref{app:additional_sim}.

\section{Conclusion}
We have presented novel fundamental results for the linear \ac{mmse} estimator and the \ac{cme} in one-bit quantized systems for \ac{gmm} distributed inputs. In addition to novel closed-form solutions for the \ac{cme}, highlighting its linearity in special cases, a new \ac{mse} inequality regarding the Gaussian and \ac{gmm} case was established, which also holds in the analyzed asymptotic regime. 
However, this inequality does not hold in the general case as the \ac{gmm} shows a more pronounced stochastic resonance effect. 
The presented results are of use for various signal processing applications.

\appendix

\subsection{Proof of Theorem \ref{theorem:lmmse}}\label{app:theorem_lmmse}
\begin{proof}
    We first observe that $\B y \sim \sum_{k=1}^K p_k \NC(\B y; \B 0, \B C_{\B y|k})$ with $\B C_{\B y|k} = \B A \B C_k \B A\h + \eta^2\eye$ due to the Gaussianity of the noise. Thus, $\B y|k \sim \NC(\B y; \B 0, \B C_{\B y|k})$, which is used in the following.
    Utilizing the law of total expectation for the definition of the Bussgang gain, cf. \cite{9307295}, yields
    \begin{align}
        \B B &= \E[Q(\B y) \B y\h] \E[\B y \B y\h]\inv 
        \\
        &= \sum_{k=1}^K p_k\E[Q(\B y) \B y\h| k] \E[\B y \B y\h]\inv.
    \end{align}
    The solution of $\E[Q(\B y) \B y\h| k] = \sqrt{\frac{2}{\pi}} \diag(\B C_{\B y|k})^{-\frac{1}{2}}\B C_{\B y|k}$ is known from the Gaussian case, cf., e.g., \cite[Sec. 9.2]{Papoulis1965ProbabilityRV}, yielding the result in \eqref{eq:Bussgang_gain_general}.
    Similarly, the cross-covariance matrix \eqref{eq:Chr} is computed as
    \begin{align}
        \B C_{\B h\B r} &= \E[\B h \B r\h] = \sum_{k=1}^K p_k \E[\B h Q(\B y)\h |k]
    \end{align}
    where $\E[\B h Q(\B y)\h] = \sqrt{\frac{2}{\pi}} \B C_k \B A\h \diag(\B C_{\B y|k})^{-\frac{1}{2}}$ is known from the Gaussian case, cf., e.g., \cite{7931630}. 
    Finally, the computation of the covariance matrix $\B C_{\B r}$ is a direct consequence of the law of total expectation, i.e.,
    \begin{align}
        \E[\B r \B r\h] = \E[Q(\B y)Q(\B y)\h] = \sum_{k=1}^K p_k\E[Q(\B y)Q(\B y)\h|k]
    \end{align}
    and the known solution for the Gaussian case, known as the \textit{arcsine law}, cf. \cite{1446497,272490}.
\end{proof}

\subsection{Proof of Theorem \ref{theorem:gmm_cme_1bit_univariate}}\label{app:theorem_cme_univariate}
\begin{proof}
We observe that $p(k|\B r) = p_k$ for all $k=1,\dots,K$ in \eqref{eq:cme_first_step} since the zero-mean \ac{gmm} is symmetric around the origin, and thus, the quantized observation is uninformative for evaluating the responsibility. The solution of 
\begin{align}
    \E[h|r,k] = \sqrt{\frac{2}{\pi}} \frac{\sigma_k^2}{\sqrt{\sigma_k^2 + \eta^2}}r
\end{align}
is known from the Gaussian case, see, e.g., \cite{fesl23cme1bit}.
\end{proof}

\subsection{Proof of Theorem \ref{theorem:mse_inequality}}\label{app:theorem_mse_inequality}
\begin{proof}
    Comparing \eqref{eq:mse_univariate} and \eqref{eq:mse_gauss}, after removing the equivalent terms and taking the square root on both sides, we need to show that 
    \begin{align}
        \sum_{k=1}^Kp_k \frac{\sigma_k^2}{\sqrt{\sigma_k^2 + \eta^2}} \leq \frac{\sum_{k=1}^K p_k\sigma_k^2}{\sqrt{\sum_{k=1}^K p_k\sigma_k^2 + \eta^2}}.
        \label{eq:proof_mse_condition}
    \end{align}
    Since the weights $p_k$ form a convex combination, \eqref{eq:proof_mse_condition} holds if $f(x) = \frac{x}{\sqrt{x + \eta^2}}$ is a concave function for all $x>0$ based on the definition of concave functions \cite[Sec. 3.1.8]{convex_opt}.
    Since 
    \begin{align}
        \frac{\partial^2}{\partial x^2} f(x) = -\frac{x + 4\eta^2}{4\sqrt{(\eta^2 + x)^5}} <0 \text{ for all } x,\eta^2 >0,
    \end{align}
    it follows that $f(x)$ is a concave function for all $x>0$. Thus, \eqref{eq:proof_mse_condition} is fulfilled, finishing the proof.
\end{proof}

\subsection{Proof of Theorem \ref{theorem:pilot_sequence}}\label{app:pilot_sequence}
\begin{proof}
    As a direct consequence of the circular symmetry of the individual Gaussians, it immediately follows that the zero-mean \ac{gmm} distribution is also circularly symmetric and thus has uniformly distributed phases \cite{340781}. Based on this, the same proof holds as in \cite[Appendix B]{fesl23cme1bit}, yielding the same \ac{mse}-optimal sequence.
\end{proof}

\subsection{Proof of Theorem \ref{theorem:cme_multiple}}\label{app:cme_multiple}
\begin{proof}
Similar to Theorem~\ref{theorem:gmm_cme_1bit_univariate}, the responsibility $p(k|\B r) = p_k$ since the pilot observations contain no amplitude information. The solution $\E[h| \B r,k]$ is given in \cite{fesl23cme1bit}. Since $\bar{\B C}_{\B y|k} = \B a\B a\h$ for all $k=1,\dots,K$, the computation of $\B C_{\B r}$ in \eqref{eq:arcsin} degenerates to the Gaussian case.
As both functions \eqref{eq:cme_linear} and \eqref{eq:cme_nonlinear} lead to the same \ac{mse} \cite{fesl23cme1bit}, they are equivalent on the discrete input domain by the uniqueness of the \ac{cme} \cite[Th. 1]{1459065}.
\end{proof}

\subsection{Additional Numerical Results}\label{app:additional_sim}

\begin{figure}[h]
    \centering
    \includegraphics{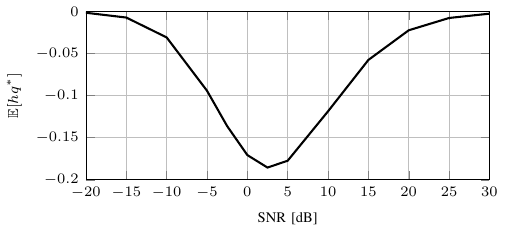}
	\caption{Correlation of the \ac{rv} of interest $h$ with the quantization noise $q$, cf. Corollary \ref{cor:corr}, for the univariate case $r = Q(h+n)$.}
    \label{fig:correlation}
\end{figure}

Fig. \ref{fig:correlation} shows the correlation $\E[hq^*]$ from \eqref{eq:correlation_quant_noise} over the \ac{snr} for the same setting as in Fig. \ref{fig:univariate_single}, which is vanishing in the low and high \ac{snr} regime in accordance with \eqref{eq:corr_vanish} and the performance loss in Fig. \ref{fig:univariate_single}.

\begin{figure}[h]
    \centering
    \includegraphics{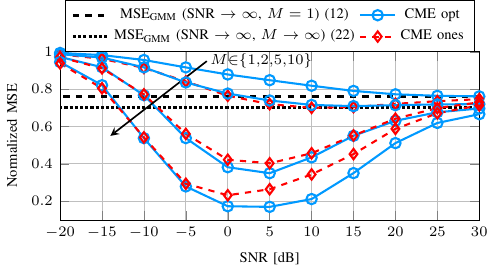}
	\caption{Comparison of the \ac{cme} with the \ac{mse}-optimal and the all-ones sequence for the case $\B r = Q(\B a h + \B n)$ with $M\in\{1,2,5,10\}$ and $\sigma_{\text{glob}}^2 = 1$.}
  \label{fig:pilots_noisy}
\end{figure}

Fig. \ref{fig:pilots_noisy} evaluates the same setting as in Fig. \ref{fig:pilots_noisy2} but compares the \ac{mse}-optimal sequence for the noiseless case derived in Theorem \ref{theorem:pilot_sequence} with the all-ones sequence $\B a = \B 1$. It can be observed that the pilot sequence from Theorem \ref{theorem:pilot_sequence} outperforms the all-ones sequence in medium to high \acp{snr}, highlighting its superiority also in the non-asymptotic \ac{snr} regime.

\balance

\bibliographystyle{IEEEtran}
\bibliography{IEEEabrv,biblio}

\end{document}